\documentclass[prd,a4paper]{revtex4}
\usepackage{epsfig}
\begin{document}

\bibliographystyle{unsrt}

\title{ Search for a promising tetraquark candidate $X(ud\bar{s}\bar{s})$ in $pn\to \Lambda\Lambda X$}

\author{Xiao-Hai Liu$^1$ and Qiang Zhao$^{1,2,3}$}

\affiliation{1) Institute of High Energy Physics, Chinese Academy
of Sciences, Beijing 100049, P.R. China}

\affiliation{2) Department of Physics, University of Surrey,
Guildford, GU2 7XH, United Kingdom}

\affiliation{3) Theoretical Physics Center for Science Facilities,
Chinese Academy of Sciences, Beijing 100049, P.R. China}

\date{\today}

\begin{abstract}

We propose to search for a tetraquark candidate
$X(ud\bar{s}\bar{s})$ in $pn\to \Lambda\Lambda
X(ud\bar{s}\bar{s})\to \Lambda\Lambda K^+K^0$ or $\Lambda\Lambda
KK^*$. The existence of tetraquark state $X(ud\bar{s}\bar{s})$ with
$J^P=0^+$, $1^-$ or $1^+$ was predicted in the literature based on
specific diquark effective degrees of freedom inside hadrons. In
order to understand the underlying dynamics for exotic hadrons, a
search for the tetraquark $X(ud\bar{s}\bar{s})$ is strongly
recommended. The proposed reaction involves two $\Lambda$
production, of which the narrow widths make it a great advantage in
the analysis of the final state missing mass spectrum. We make an
estimate of the production rate of $X(ud\bar{s}\bar{s})$ in an
effective Lagrangian theory and find that for $J^P=1^-$ the sample
events of $\sim 2200 \ nb^{-1}$ will be able to identify
$X(ud\bar{s}\bar{s})$ with five standard deviations at a width of 10
MeV to $K^+K^0$ near threshold. For $J^P=1^+$ with a width of 20 MeV
to $KK^*$, the sample events of $\sim 130 nb^{-1}$ will be needed.
Large production cross sections are expected in a kinematic region
beyond the threshold. We emphasize the advantage of low background
in this transition channel, and in the meantime caution the large
uncertainties in the present estimate due to lack of knowledge about
the $X(ud\bar{s}\bar{s})$ state. Implications for its heavy-flavored
partners $qq\bar{c}\bar{c}$ and $qq\bar{b}\bar{b}$ are briefly
discussed.

\end{abstract}

\maketitle

PACS numbers: 13.25.-k, 13.75.Ev


\vspace{1cm}

\section{Introduction}

One puzzling question in low-energy QCD is the apparent absence of
``exotic states" in hadron spectrum. Here, ``exotic states" are
referred to baryons beyond $qqq$ or mesons beyond $q\bar{q}$ which
are allowed by QCD symmetry but different from conventional
constituent quark model classifications. So far, although there
are some candidates for such exotic states, none of those has been
indisputably established in experiment~\cite{klempt-review}.

For those exotic states, if they have quantum numbers that cannot
be constructed by conventional $qqq$ or $q\bar{q}$, determination
of their exotic quantum numbers is a direct way of confirming
their existence. However, for those with conventional quantum
numbers, evidence for their existence can be contaminated by
conventional quark model configurations due to possible
configuration mixings. Their determination hence is a challenge
for both experiment and theory. In 2003, experimental evidence for
a pentaquark $\Theta^+ \ (uudd\bar{s})$ was reported by LEPS
Collaboration at SPring-8~\cite{Nakano:2003qx}, which immediately
initiated tremendous interests and activities in both experiment
and theory. With a flavor configuration of $uudd\bar{s}$, i.e.
strangeness $S=+1$, confirmation of such a state would be a direct
evidence for the existence of exotic states.

Although the final confirmation of this pentaquark state still needs
further experimental studies, it sparkled many novel ideals in the
understanding of quark-quark interactions and effective degrees of
freedom within hadrons, among which the diquark effective degrees of
freedom have attracted a lot of
attention~\cite{Jaffe:1976ig,Jaffe:2003sg,Karliner:2004sy}. We shall
not review those progresses here since there have been a lot of
detailed discussions in the literature. We only point out that as a
follow-up of the pentaquark scenario it will give rise to specific
properties of effective diquark degrees of freedom, hence will
predict existence of other multiquark exotics such as tetraquark
states due to the attractive forces between the diquark
clusters~\cite{Close:2004ip}. It is interesting to realize that such
diquark effective degrees of freedom may lead to a formation of
broad tetraquarks even though a narrow pentaquark presumably does
not exist. Studying their possible manifestations in experiment
forms our motivation of this work. In particular, we shall identify
a unique channel for a tetraquark candidate production, and
investigate possibilities of establishing or eliminating it in
future experimental measurements.

A promising  tetraquark candidate with the valence quark content
$ud\bar{s}\bar{s}$ has been broadly studied
recently~\cite{Burns:2004wy,Karliner:2004sy,KanadaEn'yo:2005zg,Vijande:2003ki,Cui:2005az,Wang:2007kb},
while some relevant earlier work has been done by Jaffe decades
ago~\cite{Jaffe:1976ig}. Burns, Close and Dudek~\cite{Burns:2004wy}
suggested that if the diquark degrees of freedom such as the ones
proposed for the pentaquark
$\Theta^{+}(uudd\bar{s})$~\cite{Jaffe:2003sg,Karliner:2004sy} exist,
there should also exist an isoscalar tetraquark $ud\bar{s}\bar{s}$
with $J^P=1^{-}$, mass about $ 1.6$ GeV, and decaying into $K^0 K^+$
with a width around $\mathcal{O}(10\sim 100)$ MeV. They used a novel
configuration to turn a ``bad diquark" into a ``good diquark".
Karliner and Lipkin~\cite{Karliner:2004sy} thought there would be an
isoscalar $ud\bar{s}\bar{s}$ meson with $J^P=0^{+}$. As it cannot
couple to $K^0 K^+$ or $KK\pi$ considering the generalized Bose
statistics and parity conservation, the lowest decay mode would be
four-body $KK\pi\pi$ channel which may result in a narrow width and
make the state detectable. In contrast, Kanada-En'yo \textit{et al.}
argued that the $ud\bar{s}\bar{s}$ meson with $J^P=1^+$ may be a
stable and low-lying state, and decay into $KK^*$ via
$S$-wave~\cite{KanadaEn'yo:2005zg}.

Motivated by the above studies, we propose to search for this
tetraquark candidate $X(ud\bar{s}\bar{s})$ in $pn\to \Lambda\Lambda
X(ud\bar{s}\bar{s})$ which is illustrated in
Fig.~\ref{feynman-diagram}. Although there are still large
uncertainties with the model predictions, the cross section may be
small and the width of $X(ud\bar{s}\bar{s})$ may be broad, the
background which contains two narrow $\Lambda$'s, is so clean that
it may be possible to extract resonance signals in the missing mass
spectrum of two $\Lambda$ final states.

\section{The model}

\begin{figure}[t]
\begin{center}
\begin{tabular}{ccc}
\includegraphics[scale=0.6]{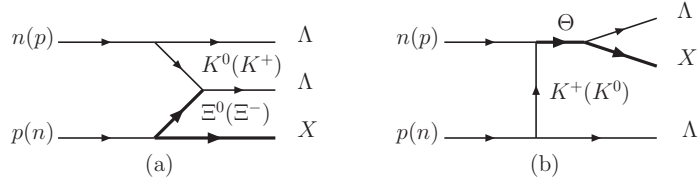}
\end{tabular}
\caption{The tetraquark candidate $X(ud\bar{s}\bar{s})$ production
via (a) $u$-channel and (b) $s$-channel transitions in $pn\to
\Lambda\Lambda X$.} \label{fig-2}
\end{center}
\end{figure}

In this section we shall study tetraquarks of $J^P=1^-$ and $1^+$ as
two most possible candidates with effective Lagrangians. In
principle, the tetraquark state can be produced via several typical
transitions as illustrated in Figs.~\ref{feynman-diagram} and
\ref{fig-2} which can be labeled by the Mandalstam variables, i.e.
$t$, $u$ and $s$-channel. In Fig.~\ref{feynman-diagram}
($t$-channel), two meson exchanges are required. In
Fig.~\ref{fig-2}(a) ($u$-channel) a meson and a $\Xi$ state will
mediate the transition while in Fig.~\ref{fig-2}(b) ($s$-channel) a
meson and a pentaquark state $\theta$ exchange are needed. Since we
have no applicable information on the $u$ and $s$-channel
transitions, we do not discuss their contributions in this work. We
shall focus on the $t$-channel transition and the details of the
model are given as follows.

\subsection{Production of $X(ud\bar{s}\bar{s})$ with $J^P=1^-$}

\begin{figure}[t]
\begin{center}
\begin{tabular}{ccc}
\includegraphics[scale=0.6]{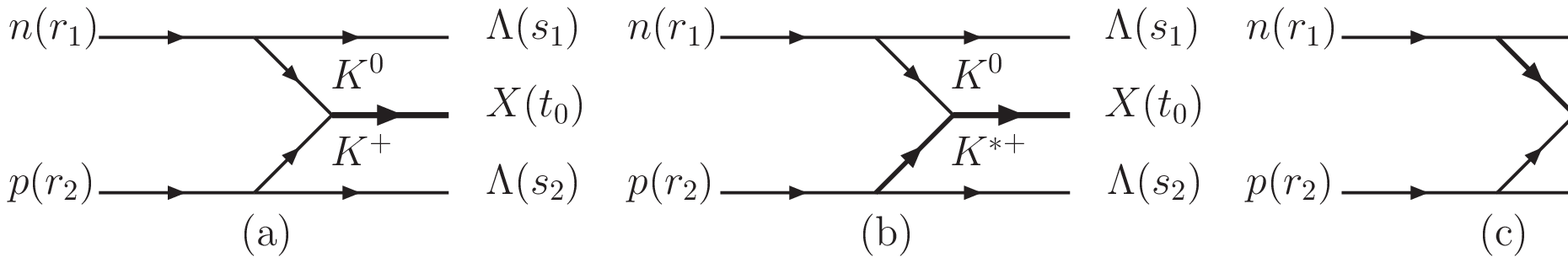}
\end{tabular}
\caption{The tetraquark candidate $X(ud\bar{s}\bar{s})$ production
via $t$-channel transitions in $pn\to \Lambda\Lambda X$.}
\label{feynman-diagram}
\end{center}
\end{figure}

For the process illustrated in Fig.~\ref{feynman-diagram}, the
following effective Lagrangians are adopted for the production of
$X(ud\bar{s}\bar{s})$ with $J^P=1^-$:
\begin{eqnarray}
\mathcal{L}_{KN\Lambda} &=&
-ig_{KN\Lambda}\overline{N}\gamma_{5}\Lambda K+H.c.  \\
\mathcal{L}_{K^{* }N\Lambda } &=& -g_{K^{\ast }N\Lambda
}\overline{N}\left(
\gamma _{\mu }\Lambda K^{\ast \mu }-\frac{\kappa _{K^{\ast }N\Lambda }}{%
2M_{N}}\sigma _{\mu \nu }\Lambda \partial ^{\nu }K^{\ast \mu
}\right) +H.c. \\
\mathcal{L}_{KKX} &=& i g_{KKX} X^\mu \left(
\partial_\mu
\overline{K^0} K^- - \overline{K^0}\partial_\mu K^- \right)+ H.c. \\
\mathcal{L}_{KK^*X} &=&
g_{KK^*X}\varepsilon^{\alpha\beta\gamma\delta}\partial_\alpha
X_\beta \partial_\gamma \bar{K^*_\delta} \bar{K} + H.c.
\end{eqnarray}
Then we adopt the following values for the $KN\Lambda$ and
$K^*N\Lambda$ couplings from Refs.~\cite{Rijken:1998yy,Stoks:1999bz}
which are also adopted by Ref.~\cite{Oh:2006hm}:
\begin{eqnarray}
&& g_{K^* N\Lambda} = -4.26, \qquad \kappa_{K^* N \Lambda} = 2.66, \nonumber \\
&& g_{KN\Lambda}^{} = - \frac{1}{\sqrt3} (1+2F) g_{\pi NN} = -13.24,
\end{eqnarray}
where $g_{KN\Lambda}^{}$ is obtained by using SU(3) flavor symmetry
relation.

The coupling constant  $g_{KKX}$  is determined by the decay width
of $X$ which is predicted by Ref.~\cite{Burns:2004wy}: $\Gamma(X\to
K^+ K^0)\ \sim \ O(10\sim 100)\ \mbox{MeV}$. With the effective
Lagrangian mentioned previously, we have:
\begin{eqnarray}
\Gamma(X\to K^+ K^0) &=&\frac{g_{KKX}^2}{48\pi}
\frac{(m_X^2-4m_K^2)^{3/2}}{ m_X^2},
\end{eqnarray}
which leads to
\begin{eqnarray}
g_{KKX}= 1.4,\ \  \mbox{when}\ \Gamma(X\to K^+ K^0)=10\ \mbox{MeV}.
\end{eqnarray}
We also assume that $\Gamma(X\to K K^*)\ \sim \ \mathcal {O}(10\sim
100)\ \mbox{MeV}$, with the similar process, we have:
\begin{eqnarray}
\Gamma(X\to K K^*) &=&\frac{g_{KK^*X}^2}{48\pi}
\frac{[(m_X^2-(m_{K^*}+m_K)^2)(m_X^2-(m_{K^*}-m_K)^2)]^{1/2}}{
2m_X^3} \nonumber \\
&\times& [(m_X^2+m_{K^*}^2-m_K^2)^2-4m_X^2 m_{K^*}^2],
\end{eqnarray}
and
\begin{eqnarray}
g_{KK^*X}= 2.6,\ \  \mbox{when}\ \Gamma(X\to K K^*)=10\ \mbox{MeV}.
\end{eqnarray}

For the vertices $KN\Lambda$ and $K^*N\Lambda$, we introduce the
covariant monopole form factor at each interaction vertex as the
intermediate meson may be off-shell,
\begin{equation}\label{formfactor}
F_M(q^2) = \frac{\Lambda^2-M_{ex}^2}{\Lambda^2-q^2},
\end{equation}
where $\Lambda=\Lambda_{KN\Lambda},\ \Lambda_{K^*N\Lambda}$ here,
and $M_{ex}$ are the corresponding masses of the exchanged mesons,
$m_K$ or $m_{K^*}$. The following commonly used values are
adopted~\cite{Oh:2006hm}:
\begin{equation}\label{ff-meson-baryon}
\Lambda_{KN\Lambda}=1.1\ \mbox{GeV},\ \ \
\Lambda_{K^*N\Lambda}=1.0\ \mbox{GeV}.
\end{equation}

So far there are no available data for the determination of the
$KKX$ and $KK^*X$ form factors. We hence introduce a dipole form
factor for the meson coupling vertices, i.e.
\begin{equation}
F_X(q^2) = \left( \frac{\Lambda_X^2-M_{ex}^2}{\Lambda_X^2-q^2}
\right)^2,
\end{equation}
where $\Lambda_X$ is the cut-off energies for $KKX$ and $KK^*X$,
and $M_{ex}$ denotes the corresponding masses of the exchanged
mesons, $m_K$ or $m_{K^*}$, respectively. The
$\Lambda_X$-dependence of the total cross section will be
illustrated later.

In Fig.~\ref{feynman-diagram}, some of the kinematic parameters
are defined as follows:
\begin{eqnarray}
q_{1} &=&r_{1}-s_{1,}\text{ \ }q_{2}=r_{2}-s_{2} \nonumber \\
w_{1} &=&r_{1}-s_{2,}\text{ \ }w_{2}=s_{1}-r_{2}.
\end{eqnarray}
The corresponding transition matrix elements are then given as
follows:
\begin{eqnarray}
T_{1fi} &=&g_{KKX}g_{KN\Lambda }^{2}\bar{u}(s_{1})\gamma _{5}u%
(r_{1})\bar{u}(s_{2})\gamma _{5}u%
(r_{2}) \frac{(q_{1}-q_{2})\cdot \epsilon
_{X}}{(q_{1}^{2}-m_{k}^{2})(q_{2}^{2}-m_{k}^{2})}F_{K}(q_1^2)F_{K}(q_2^2)F_{KKX}(q_2^2)
\nonumber\\
&-& g_{KKX}g_{KN\Lambda}^{2}\bar{u}(s_{2})\gamma _{5}u(r_{1})%
\bar{u}(s_{1})\gamma _{5}u%
(r_{2})\frac{(w_{1}-w_{2})\cdot \epsilon
_{X}}{(w_{1}^{2}-m_{k}^{2})(w_{2}^{2}-m_{k}^{2})}F_{K}(w_1^2)F_{K}(w_2^2)F_{KKX}(w_2^2),
\end{eqnarray}
\begin{eqnarray}
&&T_{2if} =  \Gamma_{KN\Lambda} \left( s_1,r_1 \right)
\Gamma_{K^*N\Lambda}^\mu\left( s_2,r_2\right)\Gamma_{KK^*X}^{\beta
\delta}\left(t_0,q_2\right) \frac{iP_{\mu
\delta}\left(q_2\right)\epsilon_\beta}{(q_1^2 - m_k^2)(q_2^2 -
m_{K^*}^2)}
F_{K}(q_1^2)F_{K^*}(q_2^2)F_{KK^*X}(q_2^2) \nonumber \\
&&- \Gamma_{KN\Lambda} \left( s_2,r_1 \right)
\Gamma_{K^*N\Lambda}^\mu\left( s_1,r_2\right)\Gamma_{KK^*X}^{\beta
\delta}\left(t_0,w_2\right) \frac{iP_{\mu
\delta}\left(w_2\right)\epsilon_\beta}{(w_1^2 - m_k^2)(w_2^2 -
m_{K^*}^2)} F_{K}(w_1^2)F_{K^*}(w_2^2)F_{KK^*X}(w_2^2) ,\nonumber \\
\end{eqnarray}
where
\begin{eqnarray}
&&\Gamma _{KN\Lambda }\left( s_{1},r_{1}\right) =-ig_{KN\Lambda }\overline{u}%
\left( s_{1}\right) \gamma _{5}u\left( r_{1}\right) \\
&&\Gamma _{K^{\ast }N\Lambda }^{\mu }(s_{2},r_{2})=g_{K^{\ast }N\Lambda }%
\overline{u}\left( s_{2}\right) \left( \gamma ^{\mu }-\frac{i\kappa
_{_{K^{\ast }N\Lambda }}\sigma ^{\mu \nu }q_{2\nu }}{2M_{N}}\right)
u\left( r_{2}\right) \\
&&\Gamma _{KK^{\ast }X}^{\beta \delta }(t_{0},q_{2})=g_{KK^{\ast
}X}\varepsilon ^{\alpha \beta \gamma \delta }t_{0\alpha }q_{2\gamma
}
\\
&& P_{\mu \delta }(q_{2})=g_{\mu \delta }-\frac{q_{2\mu }q_{2\delta }}{%
m_{K^{\ast }}^{2}} \ .
\end{eqnarray}
The transition matrix element $T_{3fi}$ for
Fig.~\ref{feynman-diagram}(c) can be obtained by making a momentum
substitution in $T_{2fi}$, i.e. $r_1\rightleftharpoons r_2$ and
$s_1 \rightleftharpoons s_2$.

\subsection{Production of $X(ud\bar{s}\bar{s})$ with $J^P=1^+$}

As studied in~\cite{KanadaEn'yo:2005zg}, axial vector meson
$ud\bar{s}\bar{s} \ (J^P=1^+)$ is a good tetraquark candidate.
Within a flux-tube model, this state might appear around $1.4\
\mbox{GeV}$ with a width of $\mathcal {O}(20\sim 80)\ \mbox{MeV}$.
We apply these quantities as an input to give an estimation of the
cross section. The process is similar to the previous section
except for the effective Lagrangian for the coupling of $KK^*X$:
\begin{equation}
\mathcal{L}_{KK^*X} = g_{KK^*X}\left(\partial^\alpha
\bar{K}^{*\beta}\partial_\alpha \bar{K}X_\beta - \partial^\alpha
\bar{K}^{*\beta}\partial_\beta \bar{K}X_\alpha \right)+H.c.
\end{equation}
The meson $X(ud\bar{s}\bar{s})$ with $J^P=1^+$ can be produced via
$KK^*$-exchange but not $KK$-exchange in this process. Thus we
only need to consider Fig.~\ref{feynman-diagram}(b) and (c). The
following is the transition amplitude:
\begin{eqnarray}
&&T_{2if} =  \Gamma_{KN\Lambda} \left( s_1,r_1 \right)
\Gamma_{K^*N\Lambda}^\mu\left( s_2,r_2\right)\Gamma_{KK^*X}^{\beta
\delta}\left(t_0,q_2\right) \frac{iP_{\mu
\delta}\left(q_2\right)\epsilon_\beta}{(q_1^2 - m_k^2)(q_2^2 -
m_{K^*}^2)}
F_{K}(q_1^2)F_{K^*}(q_2^2)F_{KK^*X}(q_2^2) \nonumber \\
&&- \Gamma_{KN\Lambda} \left( s_2,r_1 \right)
\Gamma_{K^*N\Lambda}^\mu\left( s_1,r_2\right)\Gamma_{KK^*X}^{\beta
\delta}\left(t_0,w_2\right) \frac{iP_{\mu
\delta}\left(w_2\right)\epsilon_\beta}{(w_1^2 - m_k^2)(w_2^2 -
m_{K^*}^2)} F_{K}(w_1^2)F_{K^*}(w_2^2)F_{KK^*X}(w_2^2) ,\nonumber \\
\end{eqnarray}
where
\begin{equation}
\Gamma_{KK^*X}^{\beta\delta}=g_{KK^*X}(q_1\cdot q_2 g^{\beta\delta}
-q_2^\beta q_1^\delta ),
\end{equation}
and the other terms are the same as the previous section. The
coupling constant $g_{KK^*X}$ is determined by the width:
\begin{eqnarray}
\Gamma(X\to KK^*)=g_{KK^*X}^2 \frac{|{\bf p}_{Kcm}|}{24\pi m_X^2}
\left[2(p_K\cdot p_{K^*})^2+m_{K^*}^2(m_K^2+|{\bf
p}_{Kcm}|^2)\right],
\end{eqnarray}
which leads to
\begin{eqnarray}
g_{KK^*X}&=&7.5,\ \  \mbox{when}\ \Gamma(X\to K K^*)=20\ \mbox{MeV}.
\end{eqnarray}

It should be cautioned that the adopted couplings still bare large
uncertainties due to our limited knowledge on the tetraquark states.
This will consequently bring uncertainties to the estimated
production cross sections. However, we emphasize that our strategy
here is to single out the $pn\to \Lambda\Lambda K^+ K^0$ and
$\Lambda\Lambda K K^*$ channel which are advantageous for detecting
the tetraquark states in experiment. Our calculation is to provide a
reasonable estimate of the feasibility for future experiments.

\subsection{Numerical Results}

Experimental data for $pn\to \Lambda\Lambda K^+ K^0$ and
$\Lambda\Lambda K K^*$ are not available so far. Interestingly,
there are a few experiments in 1980's on the inclusive reaction
$pp\to 2\Lambda+anything$~\cite{Bogolyubsky:1988ei} and exclusive
reaction $pp\to
2\Lambda+2K^+$~\cite{Aleshin:1982md,Panofsky:1987kj} motivated by
the search for dibaryon state formed by two $\Lambda$. The total
cross section of the inclusive reaction is about $19\pm 10\ \mu b$
with the beam energy at $E_n = 32.1$ GeV ($W= 7.87$ GeV), and that
of the exclusive reaction has an upper limit about $460\ nb$ at
$E_n= 7.8$ GeV ($W=4.05$ GeV). These data still possess large
uncertainties and need to be improved. But at this moment, they
can serve as a guidance for constraining our parameter space, and
give an estimate of the cross sections for $X$ tetraquark
productions.

Although the values for coupling constants can be determined
indirectly by other experimental data such as strangeness
productions and theoretical model prediction for $X\to K^+K^0$, we
still lack information about the choice of the cut-off energies in
the form factors. Therefore, a thorough investigation of the
parameter space is necessary. We first fix the parameters
$\Lambda_{KN\Lambda}$, $\Lambda_{K^*N\Lambda}$ as in
Eq.~(\ref{ff-meson-baryon}) and set
$\Lambda_{KKX}=\Lambda_{KK^*X}\equiv\Lambda_X=1.2$ GeV, which is a
commonly adopted value for meson-meson interaction form factors.
This allows us to obtain the total cross sections for different
$X$ states. We then examine the model sensitivities to $\Lambda_X$
by extracting the $\Lambda_X$ dependence of the total cross
sections at a given energy.

\begin{figure}[tb]
\begin{center}
\begin{tabular}{ccc}
\includegraphics[scale=0.28]{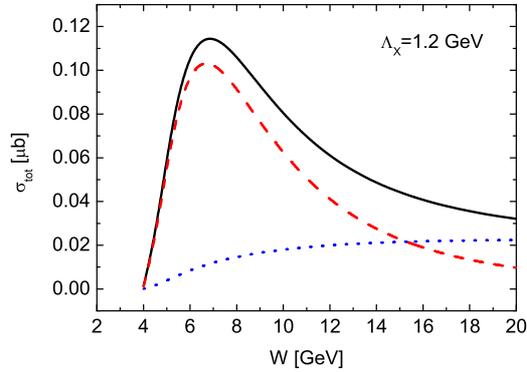}
\end{tabular}
\caption{Energy dependence of total cross sections for $pn\to
\Lambda\Lambda X$ with $J^P=1^-$ for the $X$. We have assumed
$\Gamma(X\to K^+K^0)=\Gamma(X\to KK^*)=10\
\mbox{MeV}$~\protect\cite{Burns:2004wy}. The cut-off energy is set
as $\Lambda_{KKX}=\Lambda_{KK^*X}=\Lambda_X=1.2\ \mbox{GeV}$. The
dashed line is the contribution of $KK$-exchange, dotted line is
that of $KK^*$-exchange, and solid line is the total contribution. }
\label{xection}
\end{center}
\end{figure}

\begin{figure}[tb]
\begin{center}
\begin{tabular}{ccc}
\includegraphics[scale=0.28]{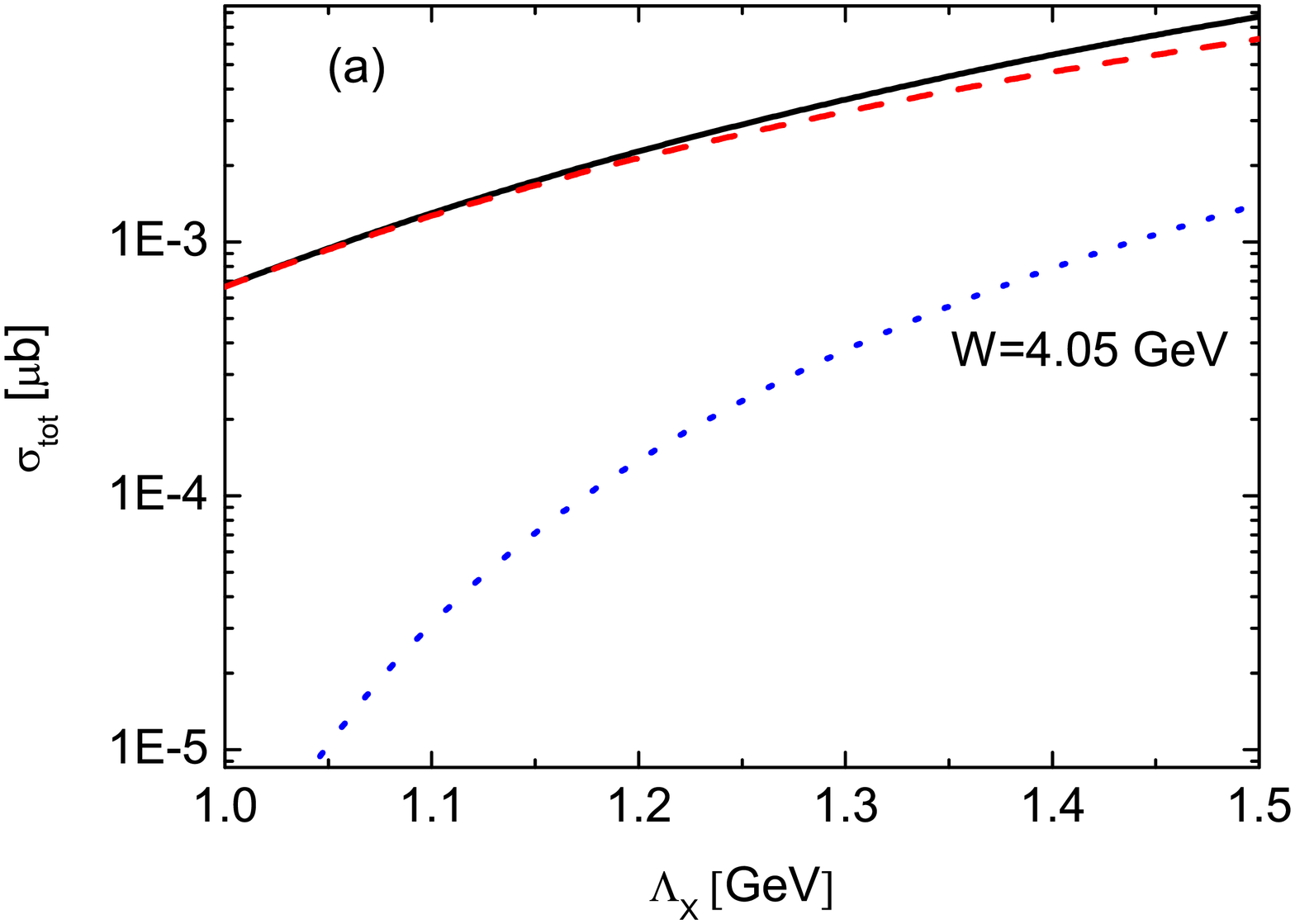}
\includegraphics[scale=0.28]{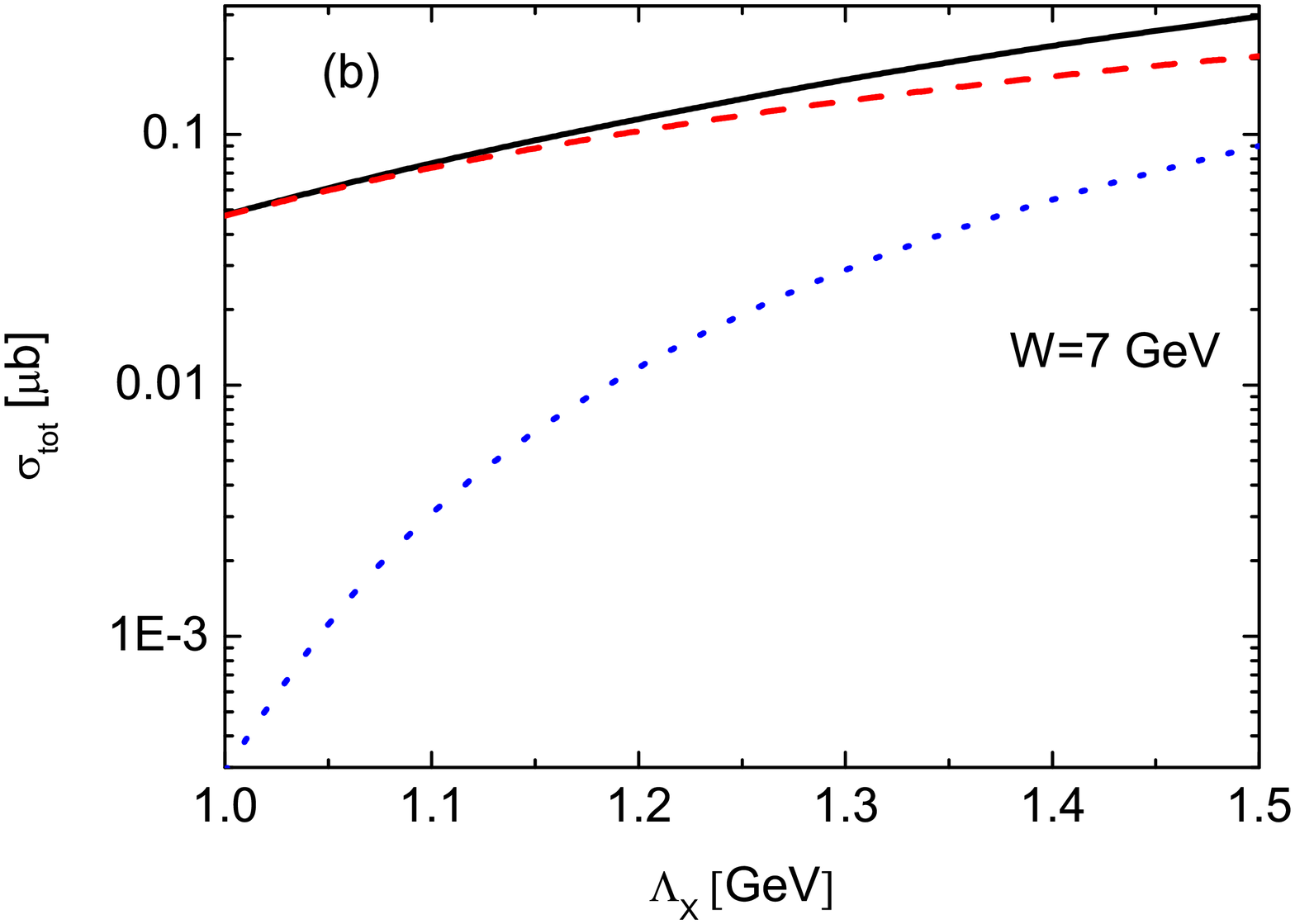}
\end{tabular}
\caption{$\Lambda_X$-dependence of total cross sections for $pn\to
\Lambda\Lambda X$ with $J^P=1^-$ at two different energies. The
notations are the same as those in Fig.~\ref{xection}. }
\label{cutoff-dep}
\end{center}
\end{figure}

In Fig.~\ref{xection}, the energy dependence of the total cross
section for $pn\to \Lambda\Lambda X$ is presented for $X$ with
$J^P=1^-$. It shows that the contribution is dominated by
$KK$-exchange, while $K K^*$-exchange is small. The total cross
section also exhibits enhancement above threshold at $W\simeq 7.0$
GeV, and then dies out with the increasing $W$. The peak value is
about 0.115 $\mu b$, which is much smaller than the cross section
for $pp\to \Lambda\Lambda +anything$~\cite{Bogolyubsky:1988ei}. This
value  is also below the upper limit of the exclusive $pp\to
2\Lambda +2 K^+$ cross section near threshold~\cite{Aleshin:1982md}.
As shown by the solid curve in Fig.~\ref{xection}, the cross section
near threshold at $E_n= 7.8$ GeV ($W=4.05$ GeV) is about 2.3 $nb$.
This value can be seen more clearly in Fig.~\ref{cutoff-dep}(a) at
$\Lambda_X=1.2$ GeV.

The sensitivity of the total cross section to the cut-off energy
$\Lambda_X$ is examined in Fig.~\ref{cutoff-dep}. The range of
$\Lambda_X=1.0\sim 1.5$ GeV is a commonly accepted one in the
literature. As shown by the solid curve, the cross section varies
between 0.05 and 0.3 $\mu b$ in terms of $\Lambda_X$, which suggests
some sensitivities to the cut-off energies. The
$\Lambda_X$-dependences of the exclusive $KK$ and $KK^*$ cross
sections are also presented by the dashed and dotted curves,
respectively, and a relatively sensitive behavior of the $KK^*$
exchange is found. However, since contributions from this transition
are rather small, the overall behavior of the cross sections is
dominated by the $KK$ exchange. Although the sensitivity to the
cut-off energy brings some uncertainties to the estimate, it does
not prevent us from drawing some preliminary conclusions on the
production rate for $X$.

\begin{figure}[tb]
\begin{center}
\begin{tabular}{ccc}
\includegraphics[scale=0.28]{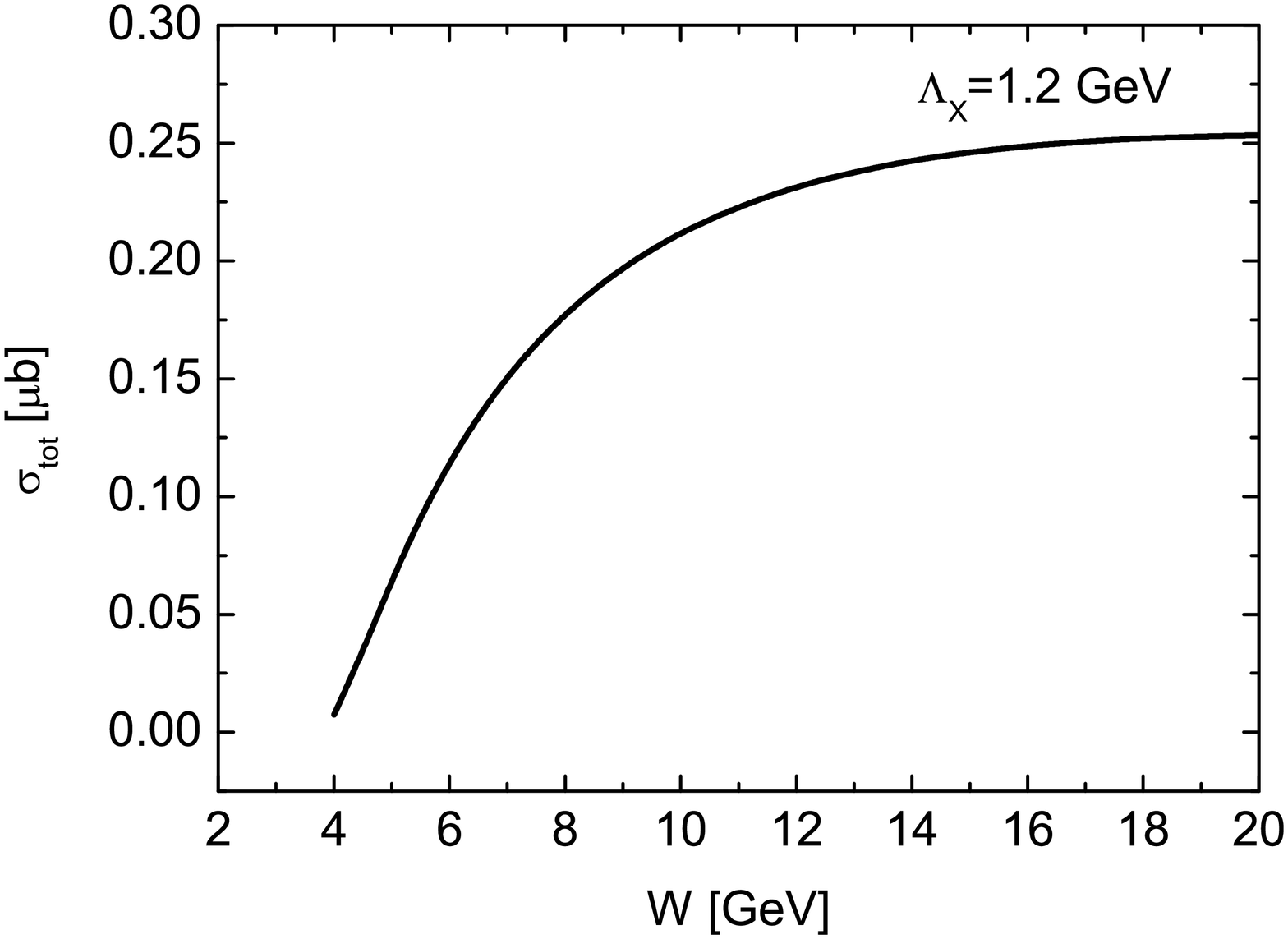}
\end{tabular}
\caption{Energy dependence of total cross sections for $pn\to
\Lambda\Lambda X$ with $J^P=1^+$ for the $X$. We have assumed
$\Gamma(X\to KK^*)=20\
\mbox{MeV}$~\protect\cite{KanadaEn'yo:2005zg}. The cut-off energy is
set as $\Lambda_{X}=1.2\ \mbox{GeV}$.  } \label{tot-axi}
\end{center}
\end{figure}

Figure~\ref{tot-axi} shows the results for $X$ production with
$J^P=1^+$. The underlying transition is via the $KK^*$ exchanges and
the cross section increases with the energies. A fast rise appears
near threshold, and then the cross section becomes rather flat. This
behavior is different from the case of $J^P=1^-$, where an obvious
enhancement appears above threshold. At $W<10$ GeV, the total cross
section is less than 0.21 $\mu b$. Then, it slowly increases to 0.25
$\mu b$ at $W=20$ GeV. Nevertheless, the cross sections for $1^+$
are relatively larger than those for $1^-$ over a wide range of $W$.
The different behavior of the near threshold cross sections between
$1^-$ and $1^+$ makes it an interesting place for looking for the
$X$ state in experiment.

\begin{figure}[tb]
\begin{center}
\begin{tabular}{ccc}
\includegraphics[scale=0.28]{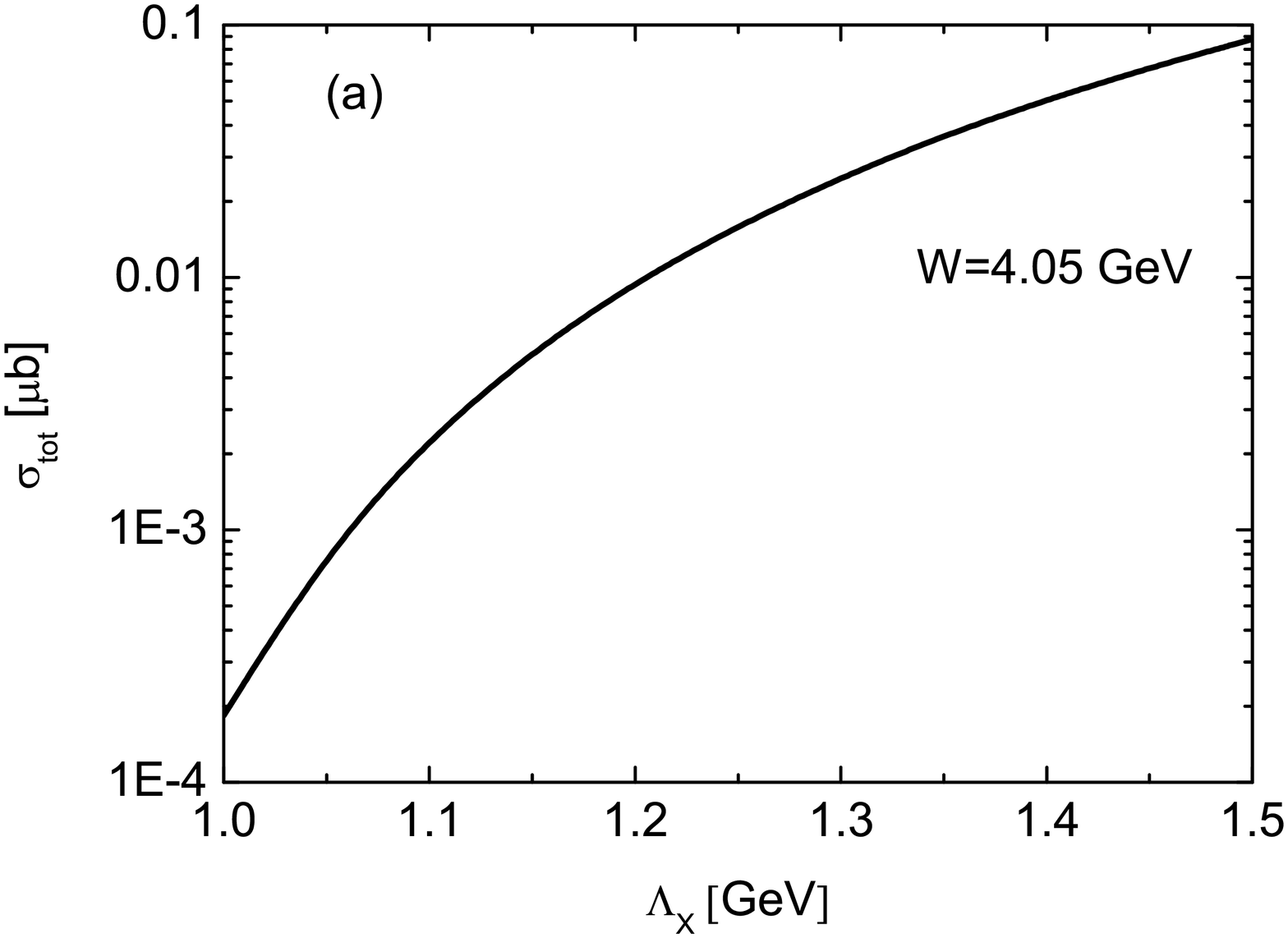}
\includegraphics[scale=0.28]{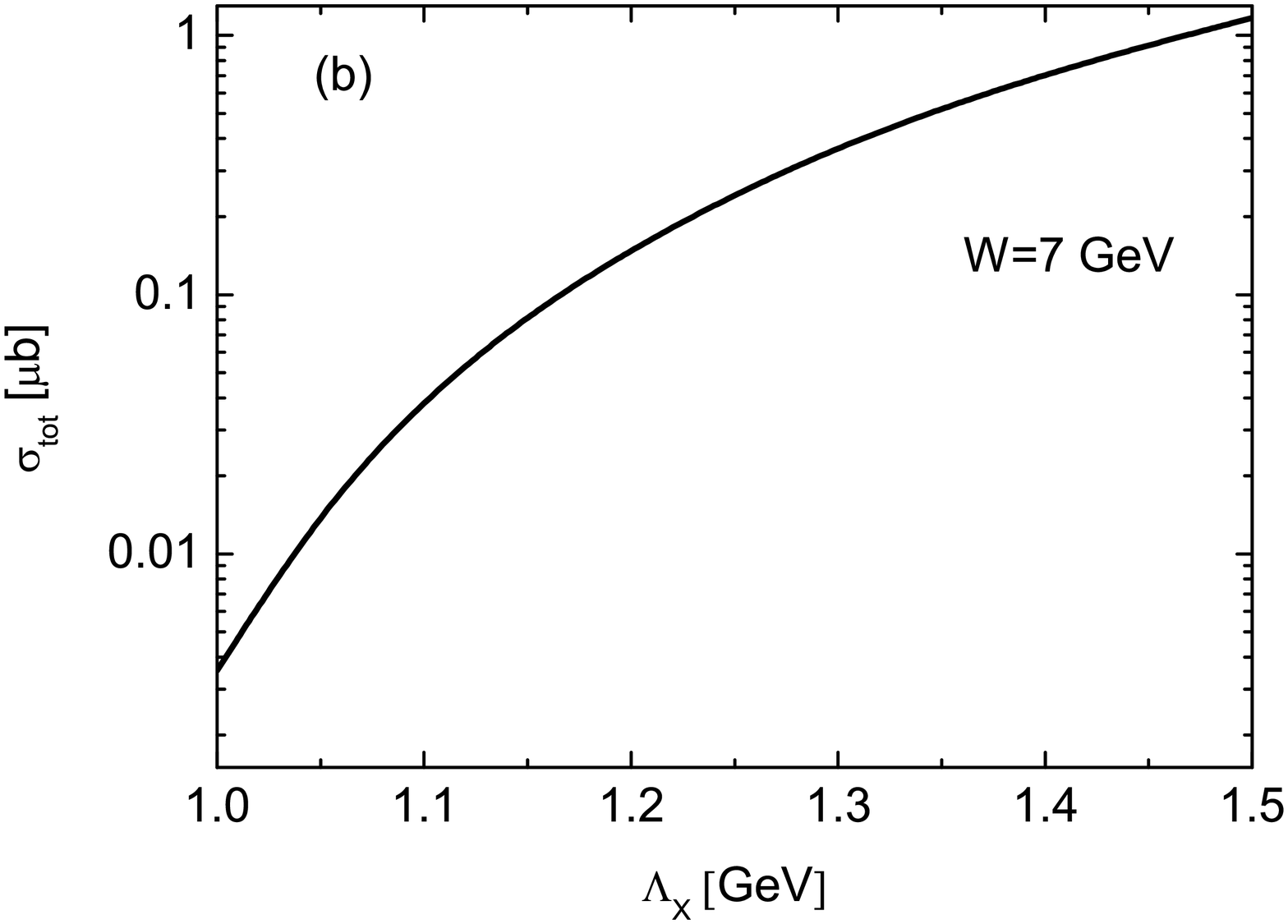}
\end{tabular}
\caption{$\Lambda_X$-dependence of total cross sections for the $X$
with $J^P=1^+$ at two different energies. } \label{cutoff-axi}
\end{center}
\end{figure}

The $\Lambda_X$-dependence of the total cross section is also
investigated and the results are displayed in Fig.~\ref{cutoff-axi}.
The cross section turns to be more sensitive to $\Lambda_X$ than the
case of $J^P=1^-$. This will increase the uncertainties of the model
predictions. However, in terms of gaining a rough idea about the
production rate of $X$ and providing a guidance of the future
experimental plan, the range of the uncertainties can still be
regarded as acceptable.

Taking the experimental data for $pp\to 2\Lambda +2 K^+$
~\cite{Aleshin:1982md} as a guidance, we can estimate the number of
sample events for establishing the $X(ud\bar{s}\bar{s})$. It is
reasonable to assume that the cross section from the background
contributions is $\sigma_{bkg} =460 \ nb$ ($W=4.05 \mbox{GeV}$),
while the signal cross section is $\sigma_X \simeq 2.3 \ nb$
($J^P=1^-$) or $\sigma_X \simeq 9.5 \ nb$ ($J^P=1^+$). With a
luminosity of $L$ and event-collecting time $t$, we require the
tetraquark signal of five standard deviations (5$\sigma$):
\begin{equation}
\frac{N_X}{\sqrt{N_{bkg}}}> 5,
\end{equation}
where $N_X$ and $N_{bkg}$ are the sample events for
$X(ud\bar{s}\bar{s})$ and the background, respectively, and they are
given by
\begin{eqnarray}
N_X &\equiv & L\times t\times \sigma_X \ , \\
N_{bkg} & \equiv & L\times t \times \sigma_{bkg} \ .
\end{eqnarray}
This leads to:
\begin{eqnarray}
&&N_{bkg}> 1.0\times 10^6\ \mbox{or}\ N_X > 5.0\times 10^3 \
\mbox{or}\
L\times t > 2174\ nb^{-1},\ \mbox{for}\ J^P=1^- \ , \\
\mbox{and} \ &&N_{bkg}> 5.8\times 10^4\ \mbox{or}\ N_X > 1.2\times
10^3 \ \ \mbox{or}\ L\times t > 127\ nb^{-1},\ \ \mbox{for}\
J^P=1^+ \ .
\end{eqnarray}

Note that the above estimations are in the near-threshold region
with $W= 4.05$ GeV instead of the kinematics with the largest cross
sections, e.g. $W\simeq 7$ GeV. The reason is that we would like to
compare our predictions with the only relevant experimental data
from Ref.~\cite{Aleshin:1982md}, of which the inclusive cross
section provide a rough check for the self-consistency of our model.
It would be more interesting to look at the energy region of $W=7\
\mbox{GeV}$ where the cross sections are predicted to have a maximum
for $J^P=1^-$ and be sizeable for $J^P=1^+$. Given the experimental
availability in the future, larger cross sections at $W=7\
\mbox{GeV}$ will make the detection of $X$ much easier there.

\section{Summary}

In this work we studied the production of possible tetraquark
candidates $X(ud\bar{s}\bar{s})$ in $pn\to \Lambda\Lambda X$.
Difficulty in the search for such an exotic state lies on the
generally-large background in its productions, while as here we
propose that $pn\to \Lambda\Lambda X(ud\bar{s}\bar{s})\to
\Lambda\Lambda K^+K^0$ and $\Lambda\Lambda KK^*$ are rather clean
and ideal for looking for its signals.

The numerical results for the cross sections were presented with a
reasonable consideration of the parameter spaces. For
$X(ud\bar{s}\bar{s})$ of $J^P=1^-$, an obvious enhancement were
observed above threshold, while for $J^P=1^+$, no enhancement was
seen. In both cases, typical values of hundreds of $n b$ were found
for the total cross section above the threshold region.

We adopted the experimental upper limit of $460\ nb$ at $E_n= 7.8$
GeV ($W=4.05$ GeV)~\cite{Aleshin:1982md}, and estimate the
signal-background rate. It shows that in order to observe the signal
of $X(ud\bar{s}\bar{s})$ at the standard deviations of $5\sigma$,
the sample events of the background must be larger than $1.0\times
10^6$ for $J^P=1^-$ (or $5.8\times 10^4$ for $J^P=1^+$), or the
product of luminosity and experiment time should be larger than
about $2174 \ nb^{-1}$ for $J^P=1^-$ (or $127\ nb^{-1}$ for
$J^P=1^+$). It should be noted that the estimates were made near
threshold where the cross sections are not the maximum. This is due
to the consideration of adopting the only available experimental
information of Ref.~\cite{Aleshin:1982md} to estimate the
signal-background rate. To search for the signal of
$X(ud\bar{s}\bar{s})$, the ideal kinematic region should be around
$W\simeq 7\ \mbox{GeV}$, where an enhancement of the cross section
was predicted for $J^P=1^-$ and the cross section also turned to be
sizeable for $J^P=1^+$. Meanwhile, we caution that although the
theoretical study of the tetraquark properties is more essentially
based on the diquark degrees of freedom, our knowledge on the
pentaquark should also have influence on the estimate of the
tetraquark decay widths. This will bring model-dependence to the
theoretical predictions. Taking into account the form factors and
couplings, the uncertainties of the results can be as large as two
orders of magnitude.

The similar experimental scheme could also be used to search for
other exotic states such as $X(qq\bar{c}\bar{c})$ in the process
$NN\to \Lambda_c\Lambda_c X$ or $X(qq\bar{b}\bar{b})$ in the
process $NN\to \Lambda_b\Lambda_b X$ and so on. Since we lack
experimental constraints on the $DN\Lambda_c$ ($BN\Lambda_b$) and
$D^*N\Lambda_c$ ($B^*N\Lambda_b$) couplings, we cannot make
quantitative estimations for those tetraquark productions. But we
stress the advantages of such a reaction process for the search
for tetraquark species of $X(qq\bar{Q}\bar{Q})$. Experimental
search for such exotics in hadron collider should be able to
provide deeper insights into the properties of strong QCD
dynamics.

\section*{Acknowledgement}

We thanks B.S. Zou for very useful discussions and comments on this
work. This work is supported, in part, by the U.K. EPSRC (Grant No.
GR/S99433/01), National Natural Science Foundation of China (Grant
No.10675131 and 10491306), and Chinese Academy of Sciences
(KJCX3-SYW-N2).

\newpage

\end{document}